\documentclass{article}
\usepackage{spconf,amsmath,graphicx}
\usepackage{bm}
\usepackage{hyperref}
\usepackage{multirow}
\usepackage{tabulary}

\PassOptionsToPackage{hyphens}{url}\usepackage{hyperref}
\hypersetup{colorlinks=true}
\newcolumntype{?}{!{\vrule width 1.5pt}}

\title{Towards lifelong learning of multilingual text-to-speech synthesis}
\name{Mu Yang$^{1}$, Shaojin Ding$^{2}$, Tianlong Chen$^{3}$, Tong Wang$^{4}$, Zhangyang Wang$^{3}$}
  
\address{$^{1}$University of Texas at Dallas, Richardson, Texas, USA \\
  $^{2}$Texas A\&M University, College Station, Texas, USA \\
  $^{3}$University of Texas at Austin, Austin, Texas, USA \\
  $^{4}$University of Science and Technology of China, Hefei, Anhui, China \\
  \normalsize $^{1}$\texttt{mu.yang@utdallas.edu}, $^{2}$\texttt{shjd@tamu.edu},\\
  \normalsize
  $^{3}$\texttt{\{tianlong.chen, atlaswang\}@utexas.edu}, $^{4}$\texttt{wtwt1997@mail.ustc.edu.cn}}
\begin{document}
\ninept
\maketitle
\begin{abstract}
This work presents a lifelong learning approach to train a multilingual Text-To-Speech (TTS) system, where each language was seen as an individual task and was learned sequentially and continually. It does not require pooled data from all languages altogether, and thus alleviates the storage and computation burden. One of the challenges of lifelong learning methods is ``catastrophic forgetting": in TTS scenario it means that model performance quickly degrades on previous languages when adapted to a new language. We approach this problem via a data-replay-based lifelong learning method. We formulate the replay process as a supervised learning problem, and propose a simple yet effective dual-sampler framework to tackle the heavily language-imbalanced training samples. Through objective and subjective evaluations, we show that this supervised learning formulation outperforms other gradient-based and regularization-based lifelong learning methods, achieving 43\% Mel-Cepstral Distortion reduction compared to a fine-tuning baseline.
\end{abstract}
\begin{keywords}
Lifelong learning, Multilingual Text-To-Speech synthesis
\end{keywords}
\section{Introduction}
\label{sec:intro}


Multilingual text-to-speech (TTS) synthesis aims to synthesize speech of different languages given corresponding input texts. Conventional Multilingual TTS systems usually require an independent model for each language~\cite{7472738,chen19f_interspeech}. More recently, end-to-end multilingual TTS system (i.e., one model for all languages) has been shown to achieve convincing performance~\cite{zhang2019learning, nekvinda20_interspeech, yang2020towards}. These systems significantly reduce the deployment complexity, which is increasingly suitable for real-world use scenarios.

Current multilingual TTS systems typically require gathering data for all target languages before launching training. In this case, it would be challenging if there were a need to support new languages. Naively fine-tuning a previously learned TTS model on new languages may not be ideal. The challenges come from several aspects in multilingual TTS scenario: different languages are diverse in linguistic contents and pronunciations \cite{zhang2019learning}. Also, Multilingual corpora usually contain unique speakers for each language \cite{park19c_interspeech}. It is thus difficult for the TTS model to preserve distinct pronunciations, language prosodies and speaker identities in previous languages while learning a new language. As a result, synthesis performance on previous languages degrades severely. From the context of lifelong learning, this is known as ``catastrophic forgetting" \cite{mccloskey1989catastrophic}.

Alternatively, one can either retrain the TTS model from scratch using both new language data and the original data, or develop co-training strategies to fine-tune the original model \cite{he2021multilingual}. Both approaches require access to new language data and full previous data that is used to train the original TTS system. Hence, although a decent performance is usually possible, it is very expensive and inefficient in terms of data storage and computation. Further, original training data may not be always available due to privacy concerns.


\begin{figure*}[htb]
    \centering
    \includegraphics[width=0.8\textwidth]{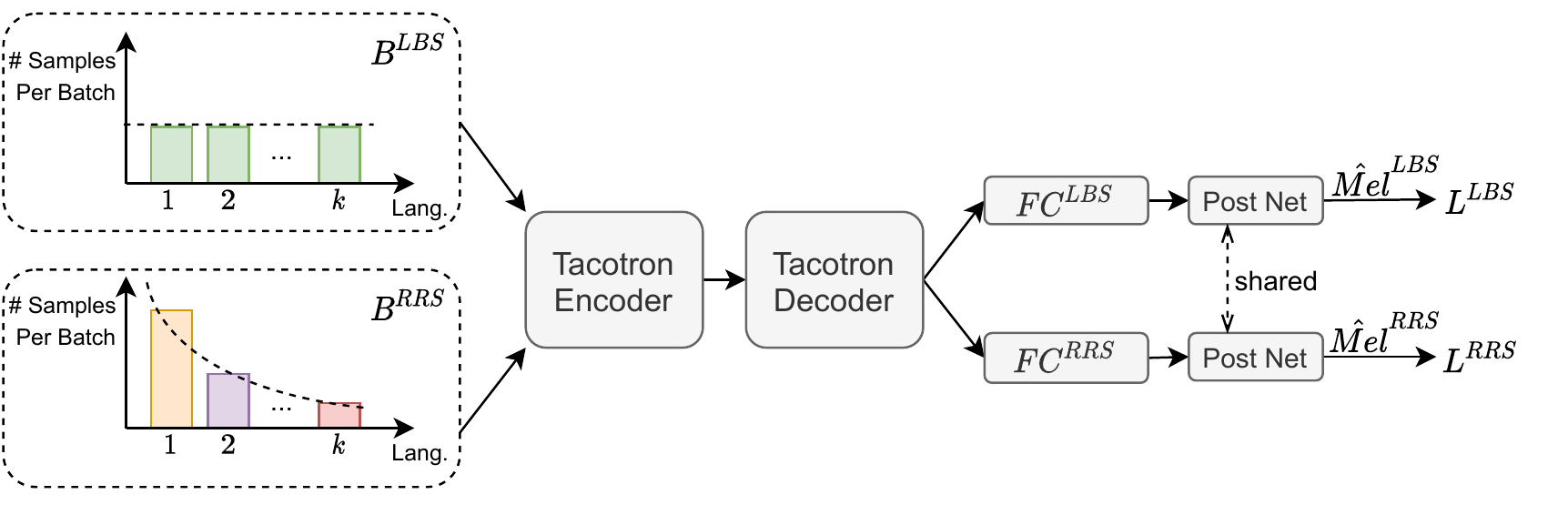}
    \vspace{-5mm}
    \caption{Proposed imbalance learning method via an auxiliary imbalance sampler. Language embedding, which is concatenated with encoder output is omitted for brevity. During inference, output is obtained from LBS branch.}
    \label{fig:proposed}
\end{figure*}

To address the problems, we for the first time propose a lifelong multilingual TTS approach, where each language is treated as an individual task and learned one-at-a-time. We approach the catastrophic forgetting problem via a data replay training scheme, where past samples are cached in a memory buffer that constraints model training on new tasks. We cast the replay process as a supervised learning problem using current language samples and the buffered \textit{small} amount of past languages samples. To address the issue of heavily language-imbalanced training samples, we propose different sampling strategies to take advantages of the full data. A novel dual-sampler framework is employed to combine benefits of both balanced sampling and random sampling. We conducted both objective and subjective evaluations on CSS10 corpus~\cite{park19c_interspeech}. Results show that the proposed method achieves 43\% Mel-Cepstral Distortion (MCD) improvement compared to the fine-tuning baseline, and it also essentially outperforms other lifelong learning baselines. 

\section{Related Work}  

\textbf{Multilingual TTS}. In recent years, several multilingual and/or cross-lingual TTS have been proposed. Some aim to establish a TTS system on low-resource languages from rich-resource languages via transfer learning \cite{lee2018learning,chen19f_interspeech}. This results in multiple TTS models for multiple languages. In contrast, another line of works train a \textit{single} TTS model for multilingual purpose \cite{zhang2019learning, nekvinda20_interspeech, yang2020towards}.
Unlike these methods that train a TTS model on multiple languages jointly, we study a more human-like multilingual TTS learning process via lifelong learning: the TTS model learns multiple languages one-at-a-time. This procedure enjoys the benefits of lower data storage burden and maintenance cost.

\noindent\textbf{Lifelong learning and applications in speech domain.} There have been extensive prior works on lifelong learning \cite{delange2021continual}. These can be grouped into 3 classes: (1) \textit{regularization}-based which consolidates previous knowledge when learning new tasks via an additional regularization term to the loss function \cite{li2017learning, kirkpatrick2017overcoming}; (2) \textit{replay}-based which employs a buffer of old samples for rehearsal/retraining or to constraint the current learning\cite{rebuffi2017icarl, belouadah2019il2m, lopez2017gradient,saha2021gradient}. (3) \textit{parameter isolation}-based which dedicate different model parameters to each task, and thus dynamically grow the network along with the tasks \cite{rusu2016progressive,xu2018reinforced,aljundi2017expert,chen2021long}. In speech domain, a handful of works have applied lifelong learning approaches on ASR \cite{sadhu20_interspeech,houston2020continual,chang21b_interspeech}, multilingual acoustic representation \cite{kessler2021continual}, fake audio detection \cite{ma2021continual} and TTS speaker adaptation \cite{hemati2021continual}. 

\section{Methods}

In this section, we describe the proposed data-replay-based lifelong multilingual TTS approach. We start by introducing the multilingual TTS framework used in this work (Sec. \ref{sec:mtts}), followed by describing our formulation of the replay methods (Sec. \ref{sec:replay_mthd}).

\subsection{Multilingual TTS framework}
\label{sec:mtts}
Our multilingual TTS model is based on Tacotron 2 \cite{shen2018natural}, an attention-based Seq2seq model. We use grapheme as encoder inputs. To account for multilingual control, one-hot encoded language embedding is concatenated with encoder output before fed into the decoder. Decoder output is projected by a fully connected layer and processed by a convolutional post net to generate Log Mel spectrogram (Fig. \ref{fig:proposed}). We follow the network architecture proposed in \cite{zhang2019learning}, except that we do not include the residual encoder, adversarial speaker classifier and speaker embedding, since we use a single-speaker corpus and do not intend to perform voice cloning. We use a WaveRNN vocoder \cite{kalchbrenner2018efficient} pre-trained on the entire CSS10 corpus to convert Mel spectrogram back to waveform.

\subsection{Proposed: Multilingual TTS through lifelong learning}
\label{sec:replay_mthd}


Ideally, to easily support new incoming languages, the multilingual TTS learner is expected to preserve a good performance on all languages, when we expose the learner into multiple languages sequentially. For this goal, the continual/lifelong learning algorithms must be able to mitigate the catastrophic forgetting on past languages, without sacrificing performance on the current language. In this study, we consider a data-replay training scheme to address this problem. Formally, let $\mathcal{D}_{k}$ denote the training set of language $k$, $\boldsymbol{\theta}_{k}^{*}$ denote the optimal model parameters to be learned on language $k$. We employ a memory buffer $\mathcal{M}_{k-1}$ that stores a limited amount of samples from past languages $1, 2, ..., k-1$. Suppose that the optimal model parameters $\boldsymbol{\theta}_{k-1}^{*}$ was learned from past languages $1, 2, ..., k-1$, our goal is to obtain $\boldsymbol{\theta}_{k}^{*}$ from $\mathcal{D}_{k}$, $\mathcal{M}_{k-1}$ and $\boldsymbol{\theta}_{k-1}^{*}$:
\begin{equation}
    \boldsymbol{\theta}_{k}^{*} \leftarrow f(\mathcal{D}_{k}^{+}, \boldsymbol{\theta}_{k-1}^{*})
\end{equation}
where $\mathcal{D}_{k}^{+}=\{\mathcal{D}_{k} \cup \mathcal{M}_{k-1}\}$, representing the merged training set of $\mathcal{D}_{k}$ and $\mathcal{M}_{k-1}$. $f$ is a learning process.


Since labeled data is buffered, we explore the feasibility of doing data replay in a supervised learning fashion:
\begin{equation}
    L(\boldsymbol{\theta}_{k}, \mathcal{D}_{k}^{+}; \boldsymbol{\theta}_{k-1}^{*}) = \frac{1}{\left|\mathcal{D}_{k}^{+}\right|} \sum_{d \in \mathcal{D}_{k}^{+}} L(\boldsymbol{\theta}_{k}, d; \boldsymbol{\theta}_{k-1}^{*})
\end{equation}
where $L$ denotes the loss function of the TTS model. $\boldsymbol{\theta}_{k-1}^{*}$ is used to initialize the training of $\boldsymbol{\theta}_{k}$. This can be formulated as an imbalanced learning problem: training samples from the new language (in $\mathcal{D}_{k}$) usually significantly outnumber buffered samples from seen languages (in $\mathcal{M}_{k-1}$), making new language the \textit{majority} language, while seen languages \textit{minority/rare} languages. Directly training a TTS model on such highly language-imbalanced samples leads to undesired outcomes - we find that the prosody of the minority languages can be ``contaminated" by that from the majority language, making them sound heavily accented (See our audio sample page). Hence, a proper re-sampling strategy for the imbalanced training data is non-trivial. We propose three sampling strategies: random sampling, weighted sampling, and sampling with a dual sampler.
\begin{table*}[ht!]
    \centering
    \begin{tabulary}{1.0\textwidth}{p{6.5em}|p{1.5em}?p{1em} p{1.5em}|p{1.6em}|p{2.6em}?p{1em} p{1em} p{1.5em}|p{1.6em}|p{2.6em}?p{1em} p{1em} p{1em} p{1.5em}|p{1.6em}|p{2.6em}}
         \hline
         \multirow{3}{6em}{Method} & \multicolumn{16}{c}{Main Training Language} \\
         \cline{2-17}
         & DE & \multicolumn{4}{c?}{NL} & \multicolumn{5}{c?}{ZH} & \multicolumn{6}{c}{JA} \\
         \cline{2-17}
         & DE & DE & NL & Avg & MCDR & DE & NL & ZH & Avg & MCDR & DE & NL & ZH & JA & Avg & MCDR  \\
         \hline
         \multirow{2}{6em}{\textbf{Lower bound}:\\ \hspace{3mm}Fine-tune} & & & & & & & & & & & & & & & & \\
         & 4.11 & 7.53 & 4.41 & 5.97 & N/A & 8.65 & 8.11 & 3.50 & 6.75 & N/A & 7.60 & 7.55 & 9.66 & 3.35 & 7.04 & N/A \\
         \multirow{2}{6em}{\textbf{Upper bound}:\\ \hspace{3mm}Joint} & & & & & & & & & & & & & & & & \\
         & 3.42 & 3.42 & 4.16 & 3.79 & 36.52\% & 3.42 & 4.16 & 3.33 & 3.64 & 46.12\% & 3.42 & 4.16 & 3.33 & 3.43 & 3.59 & 49.08\%\\
         \hline
         \multirow{2}{6em}{\textbf{Regularization}:\\ \hspace{3mm}EWC} & & & & & & & & & & & & & & & & \\
         & 4.11 & 7.40 & 4.38 & 5.89 & 1.34\% & 8.22 & 7.46 & 3.50 & 6.39 & 5.28\% & 8.18 & 7.80 & 8.44 & 3.48 & 6.98 & 0.92\% \\
         \hline
         \multirow{2}{6em}{\textbf{Replay}:\\ \hspace{3mm}GEM} & & & & & & & & & & & & & & & & \\
         & 4.11 & 4.37 & 4.56 & 4.47 & 25.21\% & 4.33 & 4.82 & 4.04 & 4.27 & 34.86\% & 4.71 & 4.87 & 4.36 & 3.68 & 4.41 & 37.43\%   \\
         \hspace{3mm}Rdm. Samp. & 4.11 & 4.39 & 4.41 & 4.40 & 26.30\% & 4.51 & 6.04 & 3.63 & 4.73 & 29.98\% & 4.77 & 5.09 & 4.41 & 3.43 & 4.43 &37.14\%\\
         \hspace{3mm}Wtd. Samp. & 4.11 & 4.67 & 4.95 & 4.81 & 19.43\% & 4.51 & 4.18 & 4.38 & 4.36 & 35.46\% & 4.90 & 4.22 & 3.57 & 3.83 & 4.13        &41.34\%\\
         \hspace{3mm}Dual Samp. & 4.11 & 4.02 & 4.30 & \textbf{4.16} & \textbf{30.37\%} & 4.15 & 4.40 & 3.89 & \textbf{4.15} & \textbf{38.52\%} & 4.56 & 4.40 & 3.85 & 3.25 & \textbf{4.02} & \textbf{42.90\%}\\
         \hline
    \end{tabulary}
    \caption{Test set MCD on seen languages throughout all training stages. Training sequence: DE-NL-ZH-JA. Column Avg denotes the average of MCDs on tested languages. MCD reduction (MCDR) is computed with respect to the lower bound performance (Fine-tune).}
    \label{tab:results}
\end{table*}

\noindent
\textbf{Random Sampling.} As the simplest sampling strategy, random sampling does not consider language imbalance. Training samples are \textit{uniformly} sampled. Hence, minority languages in $\mathcal{M}_{k-1}$ receive much less exposure than the majority language in $\mathcal{D}_{k}$. As a result, random sampling may fail to preserve performance on previous languages.

\noindent
\textbf{Weighted Sampling.} Instead of uniformly sampling training samples, a straightforward way to handle language imbalance is to sample with language-specific probabilities. Formally, let $x_{j}^{i}$ denotes the $j$th sample of language $i$, where $i=1,2,...,k$. $C_{l} = |\{ x_{j}^{i} | i=l\}|$ denotes the number of samples for language $l$. Each sample $x_{j}^{i}$ in $\mathcal{D}_{k}^{+}$ is then assigned with a weight given by the reciprocal of language-wise occurrences: $p_{j}^{i} =|\mathcal{D}_{k}^{+}| / C_{i}$, representing the chance that $x_{j}^{i}$ is to be sampled. A training batch is formed by sampling from $\mathcal{D}_{k}^{+}$ with replacement, using the assigned weights. In this way, rare languages receive higher exposure. Weighted sampling aggressively over-samples rare languages, while under-samples the majority language. This may hinder performance on current language. Also, over-fitting on rare languages may occur.

\noindent
\textbf{Dual sampler.}  On one hand, balancing language-specific samples benefits unbiased representation learning. On the other hand, simply over-sampling may result in over-fitting and affect representation learning of the majority language. Inspired by works on long-tail distribution learning \cite{liu2019large,zhang2019balance}, we propose to use a simple yet effective dual-sampler framework (Fig. \ref{fig:proposed}) to combine benefits from both balanced and imbalanced training samples.  To encourage exposure to minority languages, similar to weighted sampling, a language-balanced sampler (LBS), which samples a training batch ($B^{LBS}$) containing balanced number of samples for each language,  serves as the primary sampler to learn unbiased Seq2seq network and projection layer ($FC^{LBS}$). The regular random sampler (RRS) randomly samples a training batch ($B^{RRS}$) from the full training samples and is used as an auxiliary sampler to prevent ill-fitting of the majority languages representations. A separate projection layer $FC^{RRS}$ is utilized for RRS task, while the Tacotron Seq2seq network and post net are shared. In this way, the Seq2seq model is exposed to both sampling strategies, thus it takes advantage of full data and mitigates over-fitting on minority languages. The total training loss is computed as:
\begin{equation}
    L = \gamma L^{LBS} + \beta L^{RRS}
\end{equation}
where $L^{LBS}$ and $L^{RRS}$ are Tacotron training loss \cite{shen2018natural} from LBS and RRS task, respectively. In our experiments, we empirically set $\gamma=0.5$ and $\beta=1.0$.

\section{Experimental setup}
\label{sec:exps}
\textbf{Dataset}. We use CSS10 \cite{park19c_interspeech}, a 10-language speech corpus, with a single but different speaker for each language. We select 4 languages to form a task sequence: German (DE), Dutch (NL), Chinese (ZH), Japanese (JA). We follow the released train/validation splits from \cite{nekvinda20_interspeech}, resulting in 15.1hrs, 11.5hrs, 5.4hrs, 14.0hrs of training audio, respectively. We further split the last 20 samples from each language's validation set as their test sets. We evaluate the lifelong learning approaches using the following language order: DE-NL-ZH-JA. For replay-based methods we use a buffer size of 300 utterances, which roughly correspond to 0.6h audio. Randomly selected samples are pushed to the buffer after training on each language. When new samples need to be buffered, we randomly pop old samples to keep the buffer language-balanced throughout the entire training sequence.

\noindent \textbf{Lower/Upper bounds and baselines}. Like prior works~\cite{zhang2019learning, nekvinda20_interspeech} do, the upper bound performance can be achieved by jointly training the multilingual TTS model using full data from all 4 languages. This assumes access to the entire multilingual corpus beforehand. In contrast, the fine-tuning baseline uses only current language data to fine-tune a model trained on previous languages. This leads to aforementioned ``catastrophic forgetting" and serves as the lower bound. In addition, we also implemented two other lifelong learning algorithms and compare the proposed methods with them:
\begin{itemize}
    \item \textbf{Elastic Weight Consolidation (EWC)} \cite{kirkpatrick2017overcoming}. EWC is a regularization-based lifelong learning method.  It introduces a regularization term to penalize updates on certain model parameters that are important to previously seen languages.
    \item \textbf{Gradient Episodic Memory (GEM)} \cite{lopez2017gradient}. Similar to our proposed approach, GEM is also a replay-based method that uses a memory buffer to store past samples. When training on a new language, buffered samples constrain updates via gradient projection, such that losses on past samples do not increase.
\end{itemize}

\noindent \textbf{Model and hyper-parameters}. We train each language for 100 epochs. We adopt Adam optimizer with a initial learning rate of 0.001, decayed by half after 60 epochs. Batch size is set to 84. For all sequential learning methods, the optimal model parameters obtained from the proceeding language is used to initialize current language training.\footnote{Audio samples are available at \url{https://mu-y.github.io/speech_samples/llltts/}. Code 
is available at \url{https://github.com/Mu-Y/lll-tts}.}

\vspace{-1mm}
\section{Results}
\label{sec:results}


\vspace{-1mm}
\subsection{Objective Evaluation}
\vspace{-2mm}
\label{ssec:obj_eval}
We use Mel-Cepstral Distortion (MCD)~\cite{kubichek1993mel} between the synthesized speech and ground-truth speech as the objective evaluation metric. Test set MCDs throughout the training sequence DE-NL-ZH-JA are reported in Table \ref{tab:results}. At each training stage, we evaluate MCD on all languages seen so far (e.g. when training on ZH, we evaluate test set MCD on DE, NL and ZH). Averaged MCD on all available test languages at each stage are reported. Following \cite{chang21b_interspeech}, we compute MCD reduction (MCDR) against the lower bound baseline (\textit{Fine-tune}) to measure the recovered performance from the catastrophic forgetting brought by fine-tuning.

First, across all training stages, there are large MCD gaps between the lower bound \textit{Fine-tune} and the upper bound \textit{Joint} (e.g. 7.04 v.s. 3.59 at the final stage), demonstrating the catastrophic forgetting phenomenon. With such large gaps, we found that utterances synthesized by the \textit{Fine-tune} model are incomprehensible. Pronunciations and speaker identity information for past languages are already lost. 

Second, we can see that all our proposed data replay methods as well as \textit{EWC} and \textit{GEM} outperform the \textit{Fine-tune} baseline, with the largest improvement coming from the proposed \textit{Dual Sampling} approach. Among these methods, \textit{EWC} mitigates forgetting to a very limited extent (with a MCDR up to 5.28\% at ZH stage). This is consistent with the findings in \cite{chang21b_interspeech} which demonstrates that directly imposing constraints on model parameters via regularization may not be optimal. With \textit{Dual Sampling} strategy, the supervised-learning-based replay method outperforms the gradient-based replay method \textit{GEM}. A possible explanation is that ground truth labels of past samples may provide more useful information than gradients, and thus result in superior performance. 

Finally, we compare the three proposed sampling strategies. Overall, the proposed \textit{Dual Sampling} achieves the highest MCDR among all replay and regularization methods at all stages. As an example, at the final stage JA, \textit{Dual Sampling} achieves 42.90\% MCDR against the \textit{Fine-tune} baseline. Compared to \textit{Random Sampling}, \textit{Dual Sampling} reaches lower MCD on all seen languages so far, indicating the benefit of including a language-balanced sampler into supervised data replay. On the other hand, \textit{Weighted Sampling} is slightly better than \textit{Dual Sampling} on past languages (NL: 4.22 vs 4.40, ZH: 3.57 vs 3.85), at the cost of much higher MCD on the current language (JA: 3.83 vs 3.25). This result is consistent with our expectation: over-sampling minority languages (in this case, DE, NL, ZH in the buffer) may lead to the ill-fitting on the majority language (JA). As a result, \textit{Weighted Sampling} is worse than \textit{Dual Sampling} in terms of average MCD. At JA stage, we can also observe that, despite that \textit{Weighted Sampling} over-samples DE, MCD on DE is in fact higher than \textit{Dual Sampling}. One possible reason is that as the earliest training language, DE receives the most exposure to the TTS model with the aggressive over-sampling strategy imposed by \textit{Weighted Sampling}, making the TTS model over-fit on DE. This result show that the proposed \textit{Dual Sampling} strategy can mitigate over-fitting on minority languages. Similar general patterns can also be observed at other training stages. 


\begin{figure}[t]
    \centering
    \includegraphics[width=1.0\columnwidth]{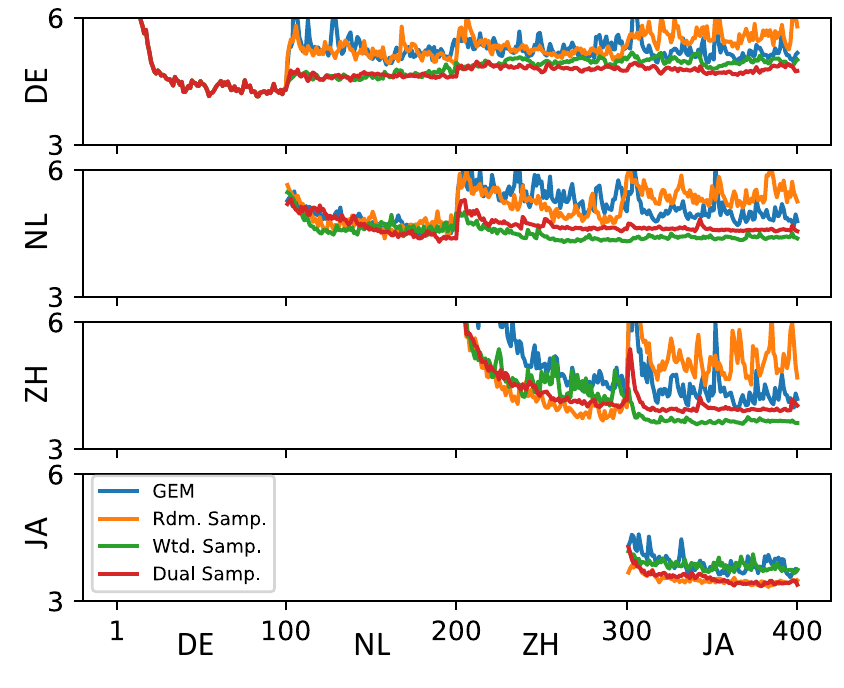}
    \vspace{-7mm}
    \caption{Dev set MCD for each language throughout the training sequence DE-NL-ZH-JA. Each language is trained for 100 epochs. Curves are plotted with a smooth factor of 0.5. }
    \label{fig:mcd_curve}
    \vspace{-3mm}
\end{figure}

\vspace{-1mm}
\subsection{Learning Curve}
\vspace{-2mm}

We show the learning curves (in terms of Dev set MCD) throughout the training sequence DE-NL-ZH-JA in Fig. \ref{fig:mcd_curve}. Due to the inferior performance, we leave out \textit{EWC} and \textit{Fine-tune} and only compare the proposed replay methods plus \textit{GEM}. At DE stage (epoch 1-100), since no past samples are available, all 4 methods are equivalent to a basic supervised learning on DE, and thus 4 curves are identical. At the next stage (NL), DE MCD increases for all methods, with larger margins for \textit{GEM} and \textit{Random Sampling}. This indicates the more severe forgetting phenomenon in \textit{GEM} and \textit{Random Sampling}. We also observe that \textit{Weighted Sampling} landed a slightly higher MCD than \textit{Dual Sampling} on current language (NL). This margin becomes larger as the training move forward (ZH at epoch 200-300, JA at epoch 300-400), demonstrating the negative effect of over-sampling past languages. On the other hand, \textit{Weighted Sampling} produces lower MCD on past languages than \textit{Dual Sampling}, due to its more intensive past languages exposure. Corroborating with results shown in Table \ref{tab:results}, we conclude that \textit{Weighted Sampling} is better capable of preserving performance on past languages but is relatively weaker on learning new languages, while \textit{Dual Sampling} yields a more balanced overall performance, as shown by the average MCD scores.

\vspace{-1mm}
\subsection{Subjective Evaluation}
\vspace{-2mm}
\begin{table}[t]
    \centering
    \resizebox{1.0\columnwidth}{!}{
    \begin{tabular}{c|c c c c}
    \hline
               & DE & NL & ZH & JA\\
    \hline
    GEM        & 2.52$\pm$0.40  & 2.02$\pm$0.26 & 2.23$\pm$0.26 & 3.69$\pm$0.20\\
    Rdm. Samp. & 2.11$\pm$0.39  & 1.78$\pm$0.36 &  2.31$\pm$0.31 & 3.64$\pm$0.30\\
    Wtd. Samp. & 3.23$\pm$0.35  &\textbf{ 3.13$\pm$0.30} &  \textbf{3.15$\pm$0.29} & 3.60$\pm$0.34\\
    Dual Samp.  & \textbf{3.28$\pm$0.39} & 3.01$\pm$0.35 & 3.02$\pm$0.28 & \textbf{3.76$\pm$0.35}\\
    \hline
    \end{tabular}}
    \caption{Test set MOS (with standard deviation) after the final training stage.}
    \vspace{-4mm}
    \label{tab:mos}
\end{table}

We conduct subjective evaluation after the entire training sequence DE-NL-ZH-JA finished. Test set utterances of all 4 languages are synthesized by the final model. We recruit participants on Amazon Mechanical Turk to score the naturalness of the synthesized audio in a scale of 1 (unnatural) to 5 (natural). Each language was evaluated by 15 native speakers. The Mean Opinion Scores (MOS) are shown in Table \ref{tab:mos}. \textit{Fine-tune} and other methods are left out because they clearly \textbf{fail} to synthesize intelligible speech. Consistent with our findings in Table \ref{tab:results} and Fig. \ref{fig:mcd_curve}, \textit{Weighted Sampling} produces more natural speech on past languages (NL, ZH), while sacrificing the performance on the final language (JA). Without a proper sampling strategy, \textit{GEM} and \textit{Random Sampling} both make ``mumbled" or heavily accented speech, which are often hard to understand. The results demonstrate the effectiveness of our proposed sampling method to mitigating the forgetting phenomenon.

\section{conclusion}
We have for the first time presented a lifelong learning approach for multilingual TTS. Our proposed supervised-learning formulation and novel sampling strategies significantly improve synthesis quality over multiple lifelong learning baselines. Nevertheless, we acknowledge that there exist more accents and/or mis-pronunciations in the synthesized utterances compared to the joint training approach. This indicates that this challenging human-like continual language learning problem for a TTS model is far from solved. In future work, we plan to investigate techniques to enhance synthesis naturalness, for example, by dedicating language-specific model parameters. Cross-lingual transfer learning may also be explored for the TTS model to adapt to new languages.


\bibliographystyle{IEEEbib}
\bibliography{refs}

\end{document}